\documentstyle[amssymb,aps]{revtex}
\begin{document}
\draft
\title{MULTIPHOTON RESONANT TRANSITIONS OF ELECTRONS IN THE LASER FIELD IN A MEDIUM }
\author{H.K. Avetissian, A.L. Khachatryan, G.F. Mkrtchian}
\address{Plasma Physics Laboratory, Department of Theoretical Physics\\
Yerevan State University\\
1, A. Manukian, 375049 Yerevan, Armenia\\
E-mail: Avetissian@ysu.am\\
Fax: (3742) 151-087 }
\maketitle

\begin{abstract}
Within the scope of the relativistic quantum theory for electron-laser
interaction in a medium and using the resonant approximation for the two
degenerated states of an electron in a monochromatic radiation field [1] a
nonperturbative solution of the Dirac equation (nonlinear over field
solution of the Hill type equation) are obtained. The multiphoton cross
sections of electrons coherent scattering on the plane monochromatic wave at
the Cherenkov resonance are obtained taking into account the specificity of
induced Cherenkov process [1, 2] and spin-laser interaction as well. In the
result of this resonant scattering the electron beam quantum modulation at
high frequencies occurs that corresponds to a quantity of an electron energy
exchange at the coherent reflection from the ''phase lattice'' of slowed
plane wave in a medium. So, we can expect to have a coherent X-ray source in
induced Cherenkov process, since such beam is a potential source of coherent
radiation itself.
\end{abstract}

\section{Introduction}

As is known the coherent interaction of electrons with a plane monochromatic
wave in a dielectric medium can be described as a resonant scattering of a
particle on the ''phase lattice'' of a traveling wave similar to the Bregg
scattering of the particle on the crystal lattice \cite{1}, \cite{2}.The
latter is obvious in the frame of reference (FR) of rest of the wave. Since
the index of refraction of a medium $n>1$ ( $n(\omega )\equiv $ $n$ as the
wave is monochromatic) in this FR there is only a static periodic magnetic
field and an elastic scattering of a particle takes place. The law of
conservation for Cherenkov process taking into account the quantum recoil
translates into the Bragg resonance condition between the de Broglie wave of
the particle and this static periodic structure. Hence, in induced Cherenkov
process the interaction resonantly connects two states of the particle which
are degenerated over the longitudinal momentum with respect to the direction
of the wave propagation: the states with longitudinal momenta $p_x$ - of the
incident particle and the states with longitudinal momenta $p_x+\ell \hbar k$
- of scattered ''Bragg'' particle, as far as the conservation law of this
process is $\left| p_x\right| =\left| p_x+\ell \hbar k\right| $ ( $\ell $-
number absorbed or radiated photons with a wave vector $k=k_x$ ). The latter
is the same as the Bragg condition of coherent elastic scattering.
Therefore, in stimulated Cherenkov process no matter how weak the wave field
is the usual perturbation theory is not applicable because of such
degeneration of the states. So, the interaction near the resonance is
necessary to describe by the secular equation \cite{1}. The latter, in
particular, reveals zone structure of the particle states in the field of
transverse electromagnetic (EM) wave in a dielectric medium \cite{1}, \cite
{2}. Note that the application of the perturbation theory ignoring the
mentioned degeneration in this process has reduced to essentially incorrect
results which have been elucidated in the paper\cite{3}.

In the present work the case of strong radiation field is considered within
the scope of the relativistic quantum theory for electron-laser interaction
in a medium. Using the resonance approximation for the above mentioned two
degenerated states in a monochromatic radiation field \cite{1} a
nonperturbative solution of the Dirac equation (nonlinear over field
solution of the Hill type equation) are obtained. The multiphoton
probabilities of free electrons coherent scattering on a strong
monochromatic wave at the Cherenkov resonance are counted, taking into
account the above mentioned specificity of induced Cherenkov process \cite{1}%
, \cite{2} and spin-laser interaction as well. In the result of this
resonant scattering the electron beam quantum modulation at high frequencies
occurs that corresponds to the electron energy exchange at the coherent
reflection from the ''phase lattice'' of slowed wave in a medium. So, we can
expect to have, in principle, a coherent X-ray source in induced Cherenkov
process, since such-quantum modulated beam is a potential source of coherent
radiation itself.

\section{NONLINEAR SOLUTION OF THE DIRAC EQUATION FOR ELECTRON IN STRONG EM
RADIATION FIELD IN\ A\ MEDIUM}

In this section we shall solve Dirac equation for a spinor particle in the
given radiation field in a medium 
\begin{equation}  \label{1}
i\frac{\partial \Psi }{\partial t}=\left[ \widehat{{\bf \alpha }}(\widehat{%
{\bf p}}-e{\bf A(\tau }))+\widehat{\beta }m\right] \Psi ,
\end{equation}
where

\begin{equation}  \label{2}
\widehat{{\bf \alpha }}=\left( 
\begin{array}{cc}
0 & {\bf \sigma } \\ 
{\bf \sigma } & 0
\end{array}
\right) ;\ \widehat{\beta }=\left( 
\begin{array}{cc}
1 & 0 \\ 
0 & -1
\end{array}
\right)
\end{equation}
are the Dirac matrices, with the ${\bf \sigma }$ Pauli matrices, $m$ and $e$
are the mass and charge of a particle respectively (here we set $\hbar =c=1$%
), $\widehat{{\bf p}}=-i\hbar {\bf \nabla }$- the operator of the
generalized momentum ${\bf A=A}(t-nx/c)$-is the vector potential of a
linearly polarized plane wave propagating in the $OX$ direction in a medium:

\begin{equation}  \label{3}
{\bf A}=\left\{ 0,A_0(\tau )\cos \omega \tau ,0\right\} ;\quad \tau =t-nx/c.
\end{equation}
We shall assume that EM wave is adiabatically switched on at $\tau =-\infty $
and switched off at $\tau =+\infty $ (${\bf A(}\tau =\mp \infty {\bf )}=0$)$%
. $

To solve the problem it is more convenient to pass to the FR of rest of the
wave ($R$ frame moving with the velocity $V=1/n$). As it is noticed, in this
FR there is only static magnetic field which will be described according to (%
\ref{3}) by the following vector potential 
\begin{equation}  \label{4}
{\bf A}_R=\left\{ 0,A_0(x^{^{\prime }})\cos k^{\prime }x^{^{\prime
}},0\right\} ,
\end{equation}
where 
\begin{equation}  \label{5}
k^{\prime }=\omega \sqrt{n^2-1}.
\end{equation}
The wave function of a particle in $R$ frame is connected with the wave
function in laboratory frame $\Lambda $ by the Lorentz transformation of the
bispinors

\begin{equation}  \label{6}
\Psi =\widehat{S}(\vartheta )\Psi _R,
\end{equation}
where 
\begin{equation}  \label{7}
\widehat{S}(\vartheta )=ch\frac \vartheta 2+\alpha _xsh\frac \vartheta 2%
~;\qquad th\vartheta =V=\frac 1n
\end{equation}
is the transformation operator. For $\Psi _R$ we have the following equation

\begin{equation}  \label{8}
i\frac{\partial \Psi _R}{\partial t^{^{\prime }}}=\left[ \widehat{{\bf %
\alpha }}(\widehat{{\bf p}^{\prime }}-e{\bf A}_R{\bf (x}^{^{\prime }}))+%
\widehat{\beta }m\right] \Psi _R.
\end{equation}
Since the interaction Hamiltonian does not depend on the time and transverse
(to the direction of the wave propagation) coordinates the eigenvalues of
the operators $\widehat{H^{\prime }}$, $\widehat{p}_y^{\prime }$, $\widehat{p%
}_z^{\prime }$ are conserved: $E^{\prime }=const$, $p_y^{\prime }=const$, $%
p_z^{\prime }=const$ and the solution of Eq.(\ref{8}) can be represented in
the form of a linear combination of free solutions of the Dirac equation
with amplitudes $a_i(x^{\prime })$ depending only on $x^{\prime }$: 
\begin{equation}  \label{9}
\Psi _R({\bf r}^{\prime }{\bf ,}t^{\prime
})=\sum\limits_{i=1}^4a_i(x^{\prime })\Psi _i^{(0)}.
\end{equation}
Here 
\[
\Psi _{1,2}^{(0)}=%
%TCIMACRO{\QOVERD( ) {E^{\prime }+m}{2E^{\prime }}}
%BeginExpansion
{E^{\prime }+m \overwithdelims() 2E^{\prime }}%
%EndExpansion
^{\frac 12}\left[ 
\begin{array}{l}
\varphi _{1,2} \\ 
\frac{\sigma _xp_x^{\prime }+\sigma _yp_y^{\prime }}{E^{\prime }+m}\varphi
_{1,2}
\end{array}
\right] \exp \left[ i(p_x^{\prime }x^{\prime }+p_y^{\prime }y^{\prime }{\bf -%
}E^{\prime }t{\bf )}\right] , 
\]
\begin{equation}  \label{10}
\Psi _{3,4}^{(0)}=%
%TCIMACRO{\QOVERD( ) {E^{\prime }+m}{2E^{\prime }} }
%BeginExpansion
{E^{\prime }+m \overwithdelims() 2E^{\prime }}%
%EndExpansion
^{\frac 12}\left[ 
\begin{array}{l}
\varphi _{1,2} \\ 
\frac{-\sigma _xp_x^{\prime }+\sigma _yp_y^{\prime }}{E^{\prime }+m}\varphi
_{1,2}
\end{array}
\right] \exp \left[ i(-p_x^{\prime }x^{\prime }+p_y^{\prime }y^{\prime }{\bf %
-}E^{\prime }t{\bf )}\right] ,
\end{equation}
where 
\begin{equation}  \label{11}
p_x^{\prime }=(E^{^{\prime }2}-p_y^{^{\prime }2}-m^2)^{\frac 12},\quad
\varphi _1=\left( 
\begin{array}{c}
1 \\ 
0
\end{array}
\right) ,\quad \varphi _2=\left( 
\begin{array}{c}
0 \\ 
1
\end{array}
\right) .
\end{equation}
The solution of Eq. (\ref{8}) in the form (\ref{9}) corresponds to an
expansion of the wave function in a complete set of the wave functions of a
electron with certain energy and transverse momentum $p_y^{\prime }$ (with
longitudinal momenta $\pm (E^{^{\prime }2}-p_y^{^{\prime }2}-m^2)^{\frac 12}$
and spin projections $S_x=\pm \frac 12$). The latter are normalized to one
particle per unit volume. Since there is symmetry with respect to the
direction ${\bf A}_R$ (the $OY$ axis) we have taken, without loss of
generality, the vector ${\bf p}^{\prime }$ in the XY plane $(p_z^{\prime
}=0) $.

Substituting Eq.(\ref{9}) into Eq.(\ref{8}) then multiplying by the
Hermitian conjugate functions and taking into account (\ref{10}) and (\ref{2}%
) we obtain a set of differential equations for the unknown functions $%
a_i(x^{\prime })$. The equations for $a_1$, $a_3$ and $a_2$, $a_4$ are
separated and for these amplitudes we have the following set of equations

\[
p_x^{\prime }\frac{da_1(x^{\prime })}{dx^{\prime }}=iep_yA_y(x^{\prime
})a_1(x^{\prime })-eA_y(x^{\prime })\left( p_x^{\prime }-ip_y^{\prime
}\right) \cdot a_3(x^{\prime })\exp (-2ip_x^{\prime }x^{\prime }), 
\]
\begin{equation}  \label{12}
p_x^{\prime }\frac{da_3(x^{\prime })}{dx^{\prime }}=-iep_yA_y(x^{\prime
})a_3(x^{\prime })-eA_y(x^{\prime })\left( p_x^{\prime }+ip_y^{\prime
}\right) \cdot a_1(x^{\prime })\exp (2ip_x^{\prime }x^{\prime }).
\end{equation}
A similar set of equations is also obtained for the amplitudes $%
a_2(x^{\prime })$ and $a_4(x^{\prime })$. For simplicity we shall assume
that before the interaction there are only electrons with specified
longitudinal momentum and spin state, i. e. 
\begin{equation}  \label{13a}
\left| a_1(-\infty )\right| ^2=1,\quad \left| a_3(+\infty )\right|
^2=0,\quad \left| a_2(-\infty )\right| ^2=0,\qquad \left| a_4(+\infty
)\right| ^2=0.
\end{equation}
From the condition of conservation of the norm we have 
\begin{equation}  \label{14}
\left| a_1(x^{\prime })\right| ^2-\left| a_3(x^{\prime })\right| ^2=const
\end{equation}
and the probability of reflection is$\left| a_{3,4}(-\infty )\right| ^2$.

The application of the following unitarian transformation

\[
a_1(x^{\prime })=b_1(x^{\prime })\exp \left( i\frac{ep_y^{\prime }}{%
p_x^{\prime }}\int_{-\infty }^{x^{\prime }}A_y(\eta )d\eta -i\frac{\vartheta
^{\prime }}2\right) 
\]
\begin{equation}
a_3(x^{\prime })=b_3(x^{\prime })\exp \left( -i\frac{ep_y^{\prime }}{%
p_x^{\prime }}\int_{-\infty }^{x^{\prime }}A_y(\eta )d\eta +i\frac{\vartheta
^{\prime }}2\right) ,  \label{15}
\end{equation}
simplifies Eq.(\ref{12}). Here $\vartheta ^{\prime }$ is the angle between
the momentum of electron and the direction of the wave propagation in the $R$
frame. The new amplitudes $b_1(x^{\prime })$ and $b_3(x^{\prime })$ satisfy
the same initial conditions: $\left| b_1(-\infty )\right| ^2=1,$ $\left|
b_3(+\infty )\right| ^2=0,$ according to (\ref{13a}).

From Eq.(\ref{12}) and Eq.(\ref{15}) for the $b_1(x^{\prime })$ and $%
b_3(x^{\prime })$ we obtain the following set of equations 
\[
\frac{db_1(x^{\prime })}{dx^{\prime }}=-f(x^{\prime })b_3(x^{\prime }) 
\]
\begin{equation}  \label{16}
\frac{db_3(x^{\prime })}{dx^{\prime }}=-f^{*}(x^{\prime })b_3(x^{\prime })
\end{equation}
where 
\begin{equation}  \label{17}
f(x^{\prime })=\frac{eA_y(t)p^{\prime }}{p_x^{\prime }}\exp \left(
-2ip_x^{\prime }x^{\prime }-i\frac{2ep_y}{p_x^{\prime }}\int_{-\infty
}^{x^{\prime }}A_y(\eta )d\eta \right) ;\quad p^{\prime }=\sqrt{%
p_y^{^{\prime }2}+p_x^{^{\prime }2}}
\end{equation}
Using the following expansion by the Bessel functions 
\[
\exp \left( -i\alpha \sin kx\right) =\sum\limits_{N=-\infty }^\infty
J_N\left( \alpha \right) \exp \left( -iNkx\right) , 
\]
we can reduce Eq. (\ref{16}) to the following form 
\[
\frac{db_1(x^{\prime })}{dx^{\prime }}=-\sum\limits_{N=-\infty }^\infty
f_N\exp \left[ -i(2p_x^{\prime }-Nk)x^{\prime }\right] b_3(x^{\prime }) 
\]
\begin{equation}  \label{18}
\frac{db_3(x^{\prime })}{dx^{\prime }}=-\sum\limits_{N=-\infty }^\infty
f_N\exp \left[ i(2p_x^{\prime }-Nk)x^{\prime }\right] b_1(x^{\prime })
\end{equation}
where

\begin{equation}  \label{19}
f_N=\frac{p^{\prime }}{2p_y^{\prime }}Nk^{\prime }J_N\left( 2\xi \frac m{%
p_x^{\prime }}\frac{p_y^{\prime }}k\right) ;\quad \quad \xi =eA/m
\end{equation}

\section{ Resonant approximation for transition amplitudes}

Because of conservation of particle energy and transverse momentum ( in $R$
frame) the real transitions in the field will occur from a $p_x^{\prime }$
state to the $-p_x^{\prime }$ one and, consequently, the probabilities of
multiphoton scattering will have a maximal values for the resonant
transitions

\begin{equation}
2p_x^{\prime }=sk^{\prime },\qquad (s=\pm 1,\pm 2...)  \label{40}
\end{equation}
The latter expresses the condition of exact resonance between the electron
de Broglie wave and the incident ''wave lattice''. In the $\Lambda $ frame
inelastic scattering takes place and the Eq.(\ref{40}) corresponds to the
well known Cherenkov conservation law

\begin{equation}
\frac{2E(1-nv\cos \vartheta )}{(n^2-1)}=s\omega  \label{41}
\end{equation}
where $\vartheta $ is the angle between the electron momentum and the wave
propagation direction in the $\Lambda $ frame (the Cherenkov angle), $v$ and 
$E$ are the electron velocity and energy.

So, we can utilize the resonant approximation keeping only resonant terms in
the Eq.(\ref{18}). Generally, in this approximation, at detuning of
resonance $\left| \delta _s\right| =\left| 2p_x^{\prime }-sk^{\prime
}\right| <<k^{\prime }$ , we have the following set of equations for the
certain $s$-photon transition amplitudes $b_1^{(s)}(x^{\prime })$ and $%
b_3^{(s)}(x^{\prime })$:

\[
\frac{db_1^{(s)}(x^{\prime })}{dx^{\prime }}=-f_s\exp \left[ -i\delta
_sx^{\prime }\right] b_3^{(s)}(x^{\prime }) 
\]
\begin{equation}  \label{20}
\frac{db_3^{(s)}(x^{\prime })}{dx^{\prime }}=-f_s\exp \left[ i\delta
_sx^{\prime }\right] b_1^{(s)}(x^{\prime })
\end{equation}
This resonant approximation is valid for the slow varying functions $%
b_1^{(s)}(x^{\prime })$ and $b_3^{(s)}(x^{\prime })$, i. e. at the following
condition

\begin{equation}  \label{21}
\left| \frac{db_{1,3}^{(s)}(x^{\prime })}{dx^{\prime }}\right| <<\left|
b_{1,3}^{(s)}(x^{\prime })\right| \cdot k^{\prime }.
\end{equation}
At first we shall solve the case of exact resonance ($\delta _s=0$).
According to the boundary conditions (\ref{14}) we have the following
solutions for the amplitudes

\begin{equation}  \label{22}
b_1^{(s)}(x^{\prime })=\frac{ch\left[ \int_{x^{\prime }}^\infty f_sd\eta
\right] }{ch\left[ \int_{-\infty }^\infty f_sd\eta \right] },\qquad
b_3^{(s)}(x^{\prime })=\frac{sh\left[ \int_{x^{\prime }}^\infty f_sd\eta
\right] }{ch\left[ \int_{-\infty }^\infty f_sd\eta \right] }
\end{equation}
and for the reflection coefficient

\begin{equation}
R^{(s)}=\left| b_3^{(s)}(-\infty )\right| ^2=th^2\left[ f_s\triangle
x^{\prime }\right]  \label{23}
\end{equation}
where $\triangle x^{\prime }$ is the coherent interaction length. The
reflection coefficient in the laboratory frame of reference is the
probability of absorption at $v<1/n$ or emission at $v>1/n$. The latter can
be obtained expressing the quantities $f_s$ and $\triangle x^{\prime }$ by
the quantities in this frame since the reflection coefficient is Lorentz
invariant. So

\begin{equation}
R^{(s)}=th^2\left[ F_s\triangle \tau \right]  \label{24}
\end{equation}
where 
\begin{equation}
F_s=\left[ \frac{(1-nv\cos \vartheta )^2}{n^2-1}+v^2\sin ^2\vartheta \right]
^{1/2}\frac{s\omega }{2v\sin \vartheta }J_s\left( \frac \xi {n^2-1}\cdot 
\frac{2mv\sin \vartheta }{\omega (1-nv\cos \vartheta )}\right) \quad \quad
\label{25}
\end{equation}
and $\triangle \tau $ for actual cases is the laser pulse duration in the $%
\Lambda $ frame. The condition of applicability of this-resonant
approximation (\ref{21}) is equivalent to the condition 
\begin{equation}
\left| F_s\right| <<\omega \text{,}  \label{26}
\end{equation}
which restricts as the intensity of the wave as well as the Cherenkov angle.
Besides, to satisfy the condition (\ref{26}) we must take into account the
very sensitivity of the parameter $F_s$ towards the argument of Bessel
function, according to Eq.(\ref{25}). For the wave intensities when $%
F_s\triangle \tau \gtrsim 1$ the reflection coefficient is in the order of
unit which can occur for the large number of photons $s>>1$ when the
argument of Bessel function $Z\sim s\gg 1$ in Eq.(\ref{25}) (according to
asymptotic behavior of Bessel function $J_s(Z)$ at $Z\simeq s\gg 1$).

Let us estimate the reflection coefficient of an electron from the laser
pulse or the most probable number of absorbed/emitted photons due to
resonance interaction in induced Cherenkov process. For the typical values
of experimental parameters of this process in the gaseous medium with the
index of refraction $n-1\sim 10^{-4}$, at the initial electron energy $E\sim
50Mev$ and Cherenkov angle $\vartheta \sim 1mrad$, during the ''Bragg
reflection'' from Neodymium laser pulse ($\omega \triangle \tau \sim 10^2$ , 
$\hbar \omega =1.17eV$ ) with an intensity $10^{10}W/cm^2$ ($\xi \sim 10^{-4}
$) electron absorbs or emits about $10^5$ photons.

For the off resonant solution, when $\delta _s\neq 0$, but $f_s^2>\delta
_s^2/4$ from Eq.(\ref{20}) for $R^{(s)}$ the following expression we obtain

\begin{equation}  \label{27}
R^{(s)}=\frac{f_s^2}{\Omega _s^2}\frac{sh^2[\Omega _s\triangle x^{\prime }]}{%
1+\frac{f_s^2}{\Omega _s^2}sh^2[\Omega _s\triangle x^{\prime }]}
\end{equation}
where $\Omega _s=\sqrt{f_s^2-\delta _s^2/4}$, which has the same behavior as
in the case of exact resonance . In opposite case when $f_s^2\preceq \delta
_s^2/4$ the reflection coefficient is a oscillating function on interaction
length.

During the coherent interaction with EM wave the quantum modulation of
particles beam density occurs too which in difference to classic one after
the interaction remains unlimitedly long (for the monochromatic beam). This
is a result of coherent superposition of particle states with various energy
and momentum due to absorbed and emitted photons in the radiation field
which is conserved after the interaction. The quantum modulated state of the
particle leads to modulation of the beam density after the interaction at
the frequency of the stimulating wave and its harmonics \cite{5}. The
density modulated particles beam can be used to generate spontaneous
superradiation \cite{6} The various radiation mechanisms of quantum
modulated beams are investigated in the works \cite{7}-\cite{9}.

In stimulated Cherenkov process the beam quantum modulation occurs if the
particles wave packet size ($\Delta x$ ) is enough large : $\Delta
x>>\lambda $ ( $\lambda $ -is the radiation wavelength ). In the opposite
case the classic modulation or bunching of the beam takes place\ (klystron
interaction scheme). From Eq. (\ref{9}) and Eq. (\ref{24}) for the electron
wave function after the reflection from the wave pulse we have the
superposition of incident and reflected electron waves (in the $R$ frame)

\begin{equation}  \label{28}
\Psi _R=a_1(-\infty )\Psi _1^{(0)}+a_3(-\infty )\Psi _3^{(0)}
\end{equation}
and in the result the probability density $\rho _R=\Psi _R^{+}\Psi _R$ is
modulated at the X-ray frequencies

\begin{equation}
\rho _R^{(s)}=1+th^2\left[ f_s\triangle x^{\prime }\right] +2\left[ 1-\frac{%
p_x^{^{\prime }2}}{E^{^{\prime }2}}\right] th\left[ f_s\triangle x^{\prime
}\right] \cos (sk^{\prime }x^{\prime }-\varphi _0)  \label{29}
\end{equation}
where 
\[
\left[ 1-\frac{p_x^{^{\prime }2}}{E^{^{\prime }2}}\right] \sin \varphi
_0=\sin \vartheta ^{\prime }. 
\]
In the laboratory frame of the reference from Eq.(\ref{7}) and Eq.(\ref{28})
we have

\begin{equation}  \label{30}
\rho ^{(s)}\simeq \frac 1{\sqrt{n^2-1}}(1+th^2\left[ F_s\triangle \tau
\right] +2th\left[ F_s\triangle \tau \right] \cos (s\omega \tau -\vartheta
^{\prime })
\end{equation}
where it is taken into account that in actual case $\left| s\omega \right|
<<E$. As is seen from Eq.(\ref{28}) the modulation depth is in the order of
unit for the intensities when $F_s\triangle \tau \sim 1$ which can be
satisfied for the moderate intensities of the laser radiation in the order
of $10^{10}W/cm^2.$

\end{document}